# Usability Evaluation of a Mobile Application in Extraordinary Environment for Extraordinary People

Moniruzzaman Bhuiyan[1], Ambreen Zaman[1] and Mahdi H. Miraz[2,3]
Institute of Information Technology, University of Dhaka, Dhaka, Bangladesh
`mb@du.ac.bd, ambreen.zaman.cse@stamforduniversity.edu.bd`
Department of Computing, Glyndŵr University, Wrexham, UK
`m.miraz@glyndwr.ac.uk`
Department of Computer Science & Software Engineering, University of Hail, KSA
`m.miraz@uoh.edu.sa`

*Abstract*- **In a contemporary world, people become dependent on electronic devices. Technologies help to clarification and structure life in many ways to meet the need of the children oriented requirements. The children suffering from disabilities (e.g. autism) has desperate needs for elucidation and structures their life. MumIES is a research based system facilitates to support and manage their living. This paper works on MumIES system to evaluate usability of the system in extraordinary environment for extraordinary people. The paper shows from the survey observation users need supporting tools to access the children's potential and challenges and to give the full support to overcome disabilities. Usability evaluation has been considered one of the key challenges to MumIES system. The paper represents analysis, design of usability studies for the extraordinary user in environment.**

I. INTRODUCTION

In contemporary world, people become dependent on electronic devices. According to the article published by The Guardian (Yeomans, 2013), children spend more time in PC, TV and mobile devices. Technologies help to clarification and structure life in many ways [1] to meet the need of the children oriented requirements. The children suffering from autism have a desperate need for elucidation and structure their life [2]. Bhuiyan et. al. [3] have introduced MumIES (Multimodal Interface based Education and Support) system which is designed for the children with special needs. The initial prototype has been developed on the Android platform and can be used on a Smartphone or other platforms for the children with special needs.

In this paper, usability testing has been taken into account for evaluation of MumIES system. Usability testing refers to an evaluation process to measure the target audience usability criteria [4]. A product or service to be usable it should have some basic criteria. These are usefulness, efficiency, effectiveness, learnability, satisfaction, accessibility. Usefulness concerns the degree to achieve user goals about a design, product or service. Efficiency measures the user's goal how much accurately it accomplished and complete within the time limit. Effectiveness refers to the product behaves in the way that users expect. Learnability measure the user's ability to learn the system. Satisfaction refers to the user's opinion about the product. Accessibility measure what makes products usable by people who have disability.

Usability testing is the portion of a large effort to improve the profitability, design decisions and minimize the frustration and errors for users [4]. To achieve the testing goal, system need to form a proper design, by gathering data to identify and measure the disability of existing product and accommodate supporting material before the product become release. Another goal is to eliminate frustration and design related problem of user's point of view. So that, user may find useful, effective, efficient and satisfactory product. For usability testing basic elements are development of research questions, use a sample for the users, represent the real life work environment, observe the point of end users, and collect the qualitative and quantitative measure for performance and preference basis.

In a software development life cycle usability testing required in every phase on the basis of researchers questionnaires, the state of the product completeness, and the time required for the solution of the problem which has been found during testing. For this, four types of test are used for product usability testing for research or software development work [4]. These are: exploratory (or formative), assessment (or summative), validation (or verification), and comparison test. A product or service design and define is the basic concern of the exploratory testing study. Its objective is to examine and analyze the effectiveness of initial design perception. In this stage critical design issues has been considered during the interaction between the user and test moderator. Whereas assessments test occur during the first or middle or after the fundamental design of the product for quantitative measurement. In validation test study, it's assure that the problem which discovered in early tests has been removed or recovered and possibility of finding errors is less than earlier testing stage. And the last test is comparison test which is conjunction with every three testing stage.

Evaluation studies are generally methodology based; can be conducted following standard guidelines for ordinary users and environment. A test plan is a standard guideline which serves as a blue print to evaluate product and user of the system [4]. A test plan consists of purpose, goals and objectives of the test; research questions; participant characteristics; method; task list; test environment, equipment and logistics; test moderator role; data to be collected and evaluation measures; report contents and presentation. The reasons for performing the test have focused on purpose, goal and objectives of the test section. Research questions required to describe the issues and queries through questionnaire to resolve the research. It is very important to determine the target user of the product by analyzing participant characteristics during testing. Method section describes the synopsis of the test plan. The task list section is consists of tasks which perform by user during testing. To perform the task specific environment and necessary equipment are required for user. The test moderate duties and responsibilities need to describe in the moderator role section. The performance and preferences data should be collected on the base of research question in the data collection section. In the report content section, it lists the points that will appear in the test report. And finally in the presentation section consists of communication results of the development team both prior to and following the report.

To analysis the usability studies of the MumIES system in perspective of Bangladesh. Some school and organizations are visited for research and investigation to conduct usability evaluation in extraordinary environment for extraordinary users.

## II. RELATED WORK

To support a wider range of people including the elder and disabled people, researchers are working actively in this area for multimodal interface [5] and learning [6] system through mobile technologies. These technologies attract children to educate themselves and other function like playing game, watching TV, making contacts, gather information etc not only this also have internet connection. On this basis, games [7] are considered for the implication of developmental and learning disabilities, special cognitive and educational needs. It [8] has been found that children who are engaged at home with various type of multimodal texts by using different media and gather skills. Multimodal text refers to the path of communication works across multiple ways. This multiple modes or ways includes language of speaking and writing, movable image or static image, audio, video, gesture, non-paper based texts etc. MumIES system is about the modes of communications in visual, text, audio, and electronic [3]. A research [9] has been focused on to enhance social skills of the children with special needs. Another research [10] has shown mobile video modeling system for the special needed children. Open-source mobile based application [2] has been taken into account to efficiently and accurately capture necessary information and behaviors. Here a behavior stands for self-attack or injury, emotionally burst which can negatively impact in the human social structure of life.

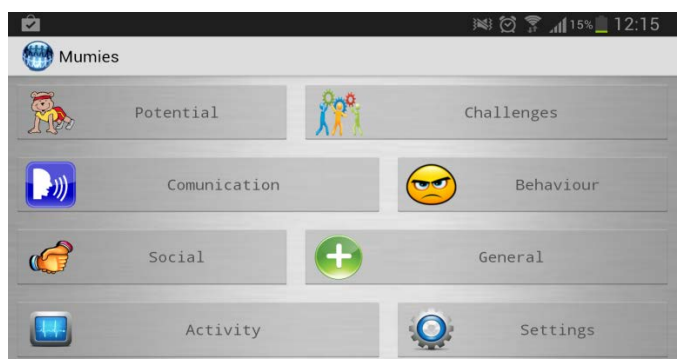

Figure 1. The interface of the MumIES system.

III. BACKGROUND STUDIES

Mobile technologies bring opportunities in learning and developing multimodal interface for needed people. A paper [1] uses this enthusiasm, for the children with ADHD *(Attention deficit hyperactivity disorder)*, who needs special care for clarity of learning and to structure their life. An ethnographic study used to identify their learning activities. In research methodology, 'etic' and 'emic' perspective has been taken into account for analysis requirement in order to generate in depth understanding of needed children's learning requirements and its educational impacts. Bhuiyan et.al [3] this paper, have introduced MumIES (developed in mobile technology) system for children with special need. A heuristic evaluation technique has been used to assess the system. For implementation of MumIES system, many schools has been visited to gather data and to prepare a specification for the system. In UK, as a part of research several cases have been studied to achieve better understanding of how the technologies impact the children. The proposed MumIES system aims is to provide self assessment and overcome the barrier and improve personal excellence, self confident and structure their life. This system takes instructions from touch, motion and speech, and also provides feedback with text, sound, light and vibration. The proposed system aimed to use Artificial Intelligence (AI) in two areas; first one is ubiquitous and multimodal activity monitoring agent. In this area intelligent interfaces act as agent tracker and keep records the activities of user. Markov decision process algorithm is used to develop the proposed intelligent scheme. The second one is to categorize multimedia libraries. For this area development user's profile based search and data mining algorithms are used. User's profile based search is made on the base of activities needs of a child and using libraries data, system suggested appropriate education and support materials. These two areas are developed in MumIES system. This system required categories are designed based on individual needs, goals and challenges. In Fig. 1 the system interface has been shown. Every general category has some necessary sub categories to have heuristic evaluation. The system develops history using graph, records activities and progresses. The researcher aim is to include interactive tools for the children, which can be developed based on their needs and activities identified by the agent.

TABLE I
THE CHARACTERISTICS OF PARTICIPANTS

| Characteristics | Desired Number of Participants |
|---|---|
| **Participant type** | |
| regular | 12 |
| backup | 4 |
| **Total number of participants** | **16** |
| **Frequency of use per day** | |
| *infrequently:* 1–5 times | 4 |
| *moderately often:* 5–12 times | 4 |
| *very often:* 13 or more times | 4 |
| **Types of user** | |
| Parents | 2 |
| Teacher/career | 8 |
| Counselor | 4 |
| Doctor | 2 |
| **Age** | |
| 21–30 | 2–3 |
| 31–40 | 4–5 |
| 41–65 | 4–5 |
| **Gender** | |
| female | 6 |
| male | 6 |

MumIES system has been managed using open source platform code.google.com to collaborate with other researchers worldwide. Code.Google is a platform in which researcher and user are able to make direct communication, exchange their thought and use codes to implement on their own devices. In this platform MumIES has been introduced in Dec, 2013. A good number of researchers (MumIES: Multi-modal Interface based Education and Support for Children with special needs) are involved in this project. Researchers from different parts of the world are contributing their findings and knowledge in the domain of the research as the authors have been working in one specific area.

IV. RESEARCH METHODOLOGY

In this research work, usability testing has been taken into account to evaluate the Smart phone based MumIES system. Usability testing criteria such as usefulness, efficiency, effectiveness, learnability, satisfaction, accessibility has been considered for proper assessment of the system. The exploratory usability studies have been designed to gather assessment data about the effectiveness of MumIES system. Usability testing plan has been taken into account to gather data to create baseline usability measurement on the basis of user perspective. The participants of the MumIES system are a forum of users, parents of children with special needs, care-worker, researchers and teachers, therapist (TABLE 1). These participants perform the main task help to accommodate data/ information to development efficient MumIES system. Questionnaires and case study has been used for data collection and to estimate the existing MumIES service. Therefore, both empirical methodologies, i.e. data about error and success rates has been measured as well as qualitative and quantitative data about participant's experience using MumIES system and other existing systems also have been considered for the research.

Few cases have been studied after observation of the participants. Some schools, resource teachers and a forum are established [12] for the children with special needs. They have been still trying to build an inclusive education system. By using MumIES, the children who need special attention within inclusive education system able to take special care and treatment. According to the case study [13], it has been found that, children who have disability had taken a special treatment for a period of time and then again he or she can keep up with her/his classmates. Some researchers [14, 15, 16, 17] has already proposed policies and planning to improve the quality of education and support system, but could not get the efficient technology based output from them. MumIES can bridge the gap in the domain. However, the intention of this paper is to introduce a MumIES system which would evaluate the usability of mobile based support system for the children with special needs in extraordinary environment for extraordinary people.

## V. Usability Evaluation in MumIES system in Bangladesh

This paper proposed to evaluate usability studies of MumIES system in extraordinary environment for the extraordinary people for the children with special need in Bangladesh. Some schools, association and foundation were visited for data collection and data analysis. Some common question has been asked to all participants. During this study, 12(twelve) individual 40 minutes usability study session conducted. Through their help it has been made possible to collect some important data for the system evaluation. The users were teachers, therapists, parents, principal of the special school, chairman of the foundation/ association, president of an association and doctor. Most of the teacher's are young and they are very much interested to have a such smart tool that guide them proper way so that they can take care of their students efficiently and effectively. They need a tool that can measure the student performance and at the same time give some suggestion for further improvement. Most of the parents are service holder. They do not get enough time to take a proper care of their disable child. They also wish to have a supporting system which help them and give them proper direction to take a good care of their children. Principal and chairman of the organization or school have the same vision as parent and teacher have, but they want more to keep track all the performance which has been measured on the basis of children development and produces a statistical outcome so that they could evaluate the performance of their school or organization. Doctor's also feel that they need a system to keep track their special patient condition and also suggest some solution through Artificial Intelligent (AI). This AI may help both the care-worker and the parents to manage and handle the autistic children when sudden difficult situation occur. All the users are facing different environments. Like as doctors are working in hospital whereas teachers, therapists or care-workers are working in a school or organization. Both are handling many children in different way. All of them need a smart tool to keep records of the children and desired to access easily to find the record(s) by simply clicking on the mobile device. It has been found that from the survey according to the user requirement MumIES system meets users' basic needs.

To meet the goal of quantitative evaluation of the research, questionnaires had been made which are as follows:

A. You are a ________
B. In perspective of Bangladesh, how successful the MumIES system will help the children with special needs?
C. Does the system/service meet the basic needs of disable children?
D. How much effectively it replace the existing supporting material or technologies to the user?

E. How well can mobile application be used, considering mobile context, mobility and slow network connection?

F. Does the system reach the users goal efficiently?

G. Is the system/service providing an appropriate balance of ease of use and ease of learning?

H. How well do users understand the symbols and icons?

I. Any further suggestion?

## VI. INTERVIEW AND OBSERVATION

The following participants were interviewed and observed on their own environment.

"X" is the Occupational Therapist, working since three years. She explains about the way they give treatment to the children. First when the child admits they observe his/her behavior and make assessment report. After that they make a goal plan for that child, which is known as Individual Education Plan (IEP).One of the goals she discussed about a child who is unable to write or color a picture. First they make sensory integration treatment in this treatment it content seven (07) sensors. These are vision sense, visual sense, auditory sense, tactile sense, muscle sense, vestibular sense, and proprioception sense. After evaluation of each sense they manage that child to finally grape a pencil in her hand and develop her understanding and concentration skill and able to write/ color picture. As she came to know about MumIES, she admits that this type of software is very helpful for both parent and teacher/carrier. Because parent wants to know their child condition and development phase which is difficult to explain sometimes. But if they have such software which not only contain data about the children strength and weakness but also produce some suggestion about how to handle and educate their children.

According to the participant "Y", she is an academic coordinator, one of the schools of autistic children in Bangladesh. They follow some specific teaching techniques which are as follows: teach own name (Who are you?); Sign follow (School, Medical, Smile, Left, Right); Written instruction; Time consult (where it is day or night or evening, what time is it? etc); Money handling(like, if you have 50/- taka on your hand and you bought a pen by 10/- taka, then calculate how many taka left?); Road Crossing; Activity of Daily Life (ADL). She also said that five (5) sense of human is not appropriate for them to sense properly. Some children do not like that somebody touch them and in some other cases child do not sense even his/her hand got hearted. They do not know how to response, even some child they do not rectify any taste of food. Inner Sense: proprioception and vestibular also taken care of. She found that most of the children are so expert on the some specific field that, if they get proper guideline, they can improve their job skill on that specific area. Like if someone is good in computer operator or like to play games in computer, in this specific area if the child gets proper direction then the child can learn how to earn from this area. This school is specifically working on to improve students' job skill on their expert area(s). Specific software has been developed for money handling system. As she came to learn about MumIES system she admit that this type of software can help them to handle those disable children to give proper guideline as they already do in paper-work and student strength and weakness can easily measure for their development.

Participant "Z" is the General Secretary of an association. He admitted that MumIES system may help to full fill the basic need which is the routine work of autistic children. As they share their experience, the autistic children have to do some regular works. There is not special treatment that they required. They need only proper care and maintain regular routine. That is why, a participant suggest to create a level basis system, like a child(x) who do not have a knowledge of the color s/he should have to practice it regularly to gain the idea of the color. But the child (y)who have already got the idea of the color knowledge s/he can color the scenery, so s/he do not need to practice color matching games or anything like that. So, here two children are different and one child is advance than another, so the care-worker who have not any idea about those children, but if s/he got a service and s/he can see the symptom of the children and can get direct idea that the children x, have different level of routine life and children y have another level of life style, this device may suggest care-worker to maintain those level through this system. According to the participant, he suggest that slow network connection may be not required if the system have multiple video's or suggestive document implicitly, so that sudden decision can be made from those useful documents. As for example, if child become very angry for the food s/he is going to take, maybe there is some AI suggestion will be appear when this type of situation arise and child may start to take food after apply those treatment which the care-worker may get from the service. Considering mobile application, he wished that this app should be in hand of all parent in very soon. He also suggest to make a link up type system, that the parent may look after whether the care-worker taking care of his/her child properly or not, he suggest to make an acknowledgement option so that parent may track that whether the routine work is working properly or not. If not than parent can give some reminder to the care-worker. According to the participant, MumIES system needs to modify in perspective of Bangladesh, because most of the parents are uneducated or know only native language. So, if the apps in native language version then the parent will use the apps properly, but still smart phone or mobile device is very expensive for some parent to buy or to maintain those devices. Because most of the parent are from rural area and it is difficult for them to buy those expensive device.

Participant "W" is a Doctor and also a chairman of an autistic children school. She and her organization are trying to develop each child according to their specific specialization field. And make them self confident and improve their skill in the right direction. They also work on their behavioral, social and communicational area to make them independent and be social. The MumIES system will help to measure the student current status and further improvement. She also hopes that the student her/himself may use for his/her own skill development purpose.

## VII. DISCUSSION AND RESULT

The result shows the importance of the usability of study, before further design, development and implementation of the MumIES system. For the usability testing research question, user's environments, and the points of end user has been collected on the basis of qualitative and quantitative measures for performance and preference data analysis. The collaboration in Google's code system provides a platform to disseminate the findings of the study with other researchers worldwide. However, through interview and observation, it has been found that system able to perform independent learning and support platform for the users. The primary findings from the qualitative investigation show significance of the study.

## VII. FUTURE WORK AND CONCLUSION

The children with special needs required sufficient resource and technology to support them and to educate those children. From the interview it has been found that care-worker and parents needs tech support for keeping track each child for the betterment of their life. It has also be found that from observation, to reach to all level of people, system need to add an option to translate the instruction into native language so that the users will use the apps properly. Another future plan is keep record in not only smart phone but other mobile devices as well to keep not only one children record but also the records of all student of a school or foundation.

ACKNOWLEDGMENT

The great acknowledge to the people and organization for their kind support. A gratitude to the IER, NFOWD, Tauri Foundation, Autism Welfare Foundation (AWF), Angels Care Foundation (ACF) and HELLO's Association of Neophyte-Students (HANS), for aided through proper resources and information regarding the children with special needs.